\documentclass[useAMS,usenatbib]{mn2e}

\usepackage[american]{babel}
\usepackage{mathrsfs,eucal,graphicx,color,amsmath,amssymb,amsfonts}
\usepackage{txfonts}
\usepackage{enumerate}
\usepackage{amsmath}

\usepackage{url}

\title[Beyond Stacking: A ML Method to Constrain Source Counts]{Beyond Stacking: A Maximum-Likelihood Method to Constrain Radio Source Counts Below the Detection Threshold}
\author[Mitchell-Wynne et al.]{Ketron~Mitchell-Wynne$^{1,2,3}$\thanks{Email: ketron@oal.ul.pt (KMW)}, M\'{a}rio G. Santos$^{3,4,5}$, Jos\'{e}~Afonso$^{1,2}$ and Matt J. Jarvis$^{4,6}$\\
$^1$Centro de Astronomia e Astrof\'{\i}sica da Universidade de Lisboa,  Observat\'{o}rio Astron\'{o}mico de Lisboa, Tapada da Ajuda, 1349-018 Lisbon, Portugal\\
$^2$Department of Physics, Faculty of Sciences, University of Lisbon, Campo Grande, 1749-016 Lisbon, Portugal\\
$^3$CENTRA, Instituto Superior T\'ecnico, Universidade T\'ecnica de Lisboa, Av. Rovisco Pais 1, 1049-001 Lisboa, Portugal\\
$^4$Department of Physics, University of the Western Cape, Bellville 7535, South Africa\\
$^5$SKA SA, 3rd Floor, The Park, Park Road, Pinelands, 7405, South Africa\\
$^6$Astrophysics, Department of Physics, University of Oxford, Keble Road, Oxford, OX1 3RH}
\date{}

\begin{document}

\date{Accepted 0000 --- 00. Received 0000 --- 00; in original form 0000 --- 00}

\pagerange{\pageref{firstpage}--\pageref{lastpage}} \pubyear{2012}

\maketitle

\label{firstpage}

\begin{abstract}
We present a statistical method based on a maximum likelihood approach to constrain the number counts of extragalactic sources below the nominal flux-density limit of continuum imaging surveys. We extract flux densities from a radio map using positional information from an auxiliary catalogue and show that we can model the number counts of this undetected population down to flux density levels well below the detection threshold of the radio survey. We demonstrate the capabilities that our method will have with future generation wide-area radio surveys by performing simulations over various sky areas. We show that it is possible to accurately constrain the number counts of the simulated distribution down to one tenth of the flux noise rms with just a sky area of 100 deg$^2$. We then test the application of our method using data from the Faint Images of the Radio Sky at Twenty-Centimeters survey (FIRST). We extract flux densities from the FIRST map, sensitive to 150~$\mu$Jy beam$^{-1}$ (1 $\sigma$), using the positional information from a catalogue in the same field, also acquired at the same frequency, sensitive to 12~$\mu$Jy beam$^{-1}$ (1 $\sigma$). Implementing our method, with known source positions, we are able to recover the right differential number counts of the noise-dominated FIRST map fluxes down to a flux density level which is one-tenth the FIRST detection threshold.

\end{abstract}


\section{Introduction}
Large astronomical surveys have become a fundamental way to understand the Universe at the largest scales. Widely known examples are the Sloan Digital Sky Survey (SDSS; \citealt{York:2000}), the Faint Images of the Radio Sky at Twenty centimeters survey (FIRST; \citealt{Becker:1994}), and the Two Micron All Sky Survey (2MASS; \citealt{Skrutskie:2007}) all of which resulted in a plethora of results applicable to essentially all fields of astronomy, demonstrating how powerful and important wide and deep surveys are, and how promising the next generation of surveys will be. Powerful statistical tools will provide an invaluable way to interpret the observations of the next generation radio telescopes, capable of observing at higher sensitivities and with higher mapping speeds, covering unprecedentedly large areas of sky.

One of the most immediate indicators to be taken from astronomical surveys are the source counts, $dN/dS$, the number of sources per flux interval per sky area. \citet{Longair:1966} first showed that cosmological evolution can be studied from such a basic quantity, and since then it has been shown that the source counts can be used to estimate details about source population and galaxy evolution (e.g. \citealt{Rowan:1970}; \citealt{Condon:1984}, and references therein).

While source counts can appear to make full use of astronomical images, the lower flux density level at which the counts can be calculated is limited by both instrumental noise from the detectors and Poisson noise from the sky. However, using the statistical properties of the noise itself, one can estimate the characteristics of the sources that lie below the detection limit. A relatively mundane example is the technique of stacking \cite[e.g.][]{Dunne2009,Karim2011}, where a collection of sources, in spite of being undetected, can have their average flux estimated by summing the signals of the respective observations while decreasing the non-pathological noise levels. The final flux estimate is frequently used as an approximate measure of the statistical properties of a well defined sample of sources, known to exist from other wavelengths. 

Over the past decade many authors have used stacking techniques to extract the average properties of populations of extragalactic sources selected in a different waveband. This has allowed the study of objects well below the usual 3--5$\sigma$ detection threshold of a given survey, providing information on star-formation rates and/or accretion activity inaccessible to standard flux-limited source catalogues. However, these stacking experiments also mean that information is lost on the distribution of sources below the flux-density limit of a given survey. A more general approach can be attempted by analysing the statistical characteristics of the noise itself, and conclusions can be drawn about the shape of the source counts in the flux density regime at or below the noise level. The method outlined in this paper can thus be used on the wealth of existing and forthcoming data at all wavelengths to extract real physical understanding of the sub-threshold source populations.

There are currently a number of methods to estimate faint radio source counts in the low flux density regime, down to $\mu$Jy levels. In the presence of confusion noise, \citet{Scheuer:1957} first showed that using a $P(D)$ analysis allows one to statistically measure the source counts from a confusion-limited survey. This $P(D)$ method has recently been used to measure number counts down to 1 $\mu$Jy at 3 GHz (\citealt{Condon:2012}). With the advent of more sensitive and higher resolution radio telescopes, the level of confusion noise drops significantly, so the extra information from sources below the detection limit will be usually dominated by instrumental noise instead of confusion noise. In this paper we assume surveys which are not limited by confusion, so a $P(D)$ analysis would not be necessary.

\citet{Crawford:1970} (see also \citealt{Murdoch:1973} and \citealt{Hogg1998}) use a maximum likelihood method to constrain faint galaxy counts at levels above the confusion noise, and has since been implemented a number of times at sub-millimetre (\citealt{Vieira2010}; \citealt{Austermann2009}) and radio wavelengths (\citealt{Fomalont:1993}; \citealt{Windhorst:1993}; \citealt{Fomalont:2002}). Although this is a robust method, the drawback is that it is to be used explicitly on sources with flux densities above some detection limit, typically 5 $\sigma$. \citet{Fomalont:1988} reports on a method which constrains radio source counts by analyzing the variances in radio maps at flux densities below the detection threshold, which allows one to measure counts down to 1 $\sigma$ (\citealt{Fomalont:1993}); although powerful, we are interested in going to even deeper flux density levels, making use of the upcoming availability of catalogues with several millions of positions where radio galaxies are known to exist.

In this paper we develop a technique based on a Maximum Likelihood approach that is a hybrid between a stacking analysis and a pure background noise analysis. The method presented here is similar to stacking in that we use positions from an external catalogue to extract fluxes from a noisy map, but it is superior to the stacking technique in that we are able to make more detailed statements about the undetected population--namely, we are able to constrain their differential number counts. Putting it simply, the combined measured fluxes from the noisy map at positions where we know that galaxies exist should provide more information than just the average, and we aim to extract that information with this new method. However, we explicitly note here that while no general remarks about the overall source population present below the detection level can be immediately drawn, it offers the unique opportunity to build partial source counts for specific populations, using previously known source positions. For example, given a catalog obtained from observations taken at ultra-violet (UV) wavelengths, one can extract fluxes from a noisy radio map at the UV detected positions, apply the method presented here, and obtain the number count contribution of UV sources at radio wavelengths. In this way, our method cannot be compared on a one-to-one basis to the $P(D)$ method, nor can it be interpreted to return the intrinsic number counts of the population in question (e.g radio in the previous example). On the other hand, as one increases the multi wavelength coverage of a given sample, it is hoped that a clear picture of the global distribution can be achieved.
Considering the new generation of deep, ultra-wide-field radio surveys which will be performed over the coming years, we develop this technique with radio surveys in mind, and exemplify its use with current state-of-the-art observations. A robust implementation of our method requires very large area surveys, e.g. with an expected large number of undetected galaxies in the images and multi-wavelength data capable of providing catalogues with large numbers of these undetected galaxies. This makes future large radio continuum surveys an ideal candidate to apply this statistical analysis.
 
The Square Kilometer Array (SKA) Phase 1 mid-frequency array will have an angular resolution of $\sim 0.2$\,arcsec and a continuum sensitivity of about 0.7 $\mu$Jy hr$^{1/2}$ \citep{Dewdney:2013}, allowing to reach $\sim 0.5\ \mu$Jy rms sensitivity over most of the visible sky and therefore, detect millions of faint galaxies over large areas of the sky \cite[see e.g.][]{JarvisRawlings2004,Wilman2008,Wilman2010}. However, due to the time table for SKA, we should also consider more near-term surveys, which are scheduled to be completed before the end of the decade.

On the shortest timescale, the Low-frequency Array \citep[LOFAR; ][]{vanHaarlem2013} will conduct a tiered survey at low radio frequencies (30--200\,MHz) reaching $\mu$Jy rms flux-densities for the first time at such low frequencies. The Karoo Array Telescope (MeerKAT; \citealt{Booth:2009}) will be able to perform observations at $\sim 1$~GHz from 2016. The MeerKAT continuum survey \citep[see e.g.][]{Jarvis2012} will conduct a two-tier survey to flux-density limits of 0.1\,$\mu$Jy and 1\,$\mu$Jy rms over $\sim 2$ and $\sim 30$~square degrees respectively.

Two additional surveys, which are expected to come online within the next few years, will image upwards of millions of sources at cm wavelengths, with rms levels of $\sim$~10~$\mu$Jy~beam$^{-1}$. The Evolutionary Map of the Universe survey (EMU; \citealt{Norris:2011}) will be carried out using the Australian Square Kilometer Array Pathfinder telescope (ASKAP; \citealt{Johnston:2008}) and is predicted to detect up to 70 million galaxies in the Southern sky on images reaching a rms of $\sim$~10~$\mu$Jy. With the installation of the APERTIF receiver \citep{Oosterloo:2009} on the Westerbrook Synthesis Radio Telescope (WSRT), the Westerbork Observations of the Deep Apertif Northern sky survey (WODAN) will detect millions of galaxies in the Northern hemisphere at 1.4~GHz with similar depth and resolution \citep{Rotter:2011}.

All the above instruments will have arc second resolutions so that confusion noise (given by the integrated contribution from undetected sources, \citealt{Condon:2009}) and ``natural'' confusion (when there is a high chance of source overlap, \citealt{Condon:2012}) are expected to be at much lower flux levels than the corresponding instrumental noise, creating ideal conditions for the application of our method. Based on the simulations of \citealt{Wilman2008,Wilman2010}, which agree with the latest results on the faint source counts (e.g. Simpson et al. 2012), EMU with a resolution of $\sim10''$ has around 70 beams per source at the 10~$\mu$Jy rms of the survey. With a resolution of 3~acrcsec at the depth of the proposed MeerKAT-MIGHTEE survey of 1~$\mu$Jy rms, we expect 30 beams per source. Given that radio sources at these flux-density levels tend to be compact ($<1''$) \citep{Mux:2005}, we will again be in the realms of instrumental confusion rather than natural source confusion.

\section{Statistical description}
The objective in this analysis is to constrain the source number density $dN/dS$, for a population of galaxies, below the detection threshold (usually 3 to 5 times the rms noise) for a given radio survey\footnote{As mentioned before, this analysis can be extended to other wavelengths}. We can take advantage of the knowledge of the positions of sources, undetected in the radio, from other (e.g., optical, near-infrared) surveys. 
By measuring the radio flux densities at these known positions, we will obtain $N$ noise-dominated flux density estimates ($S_m$) between some minimum $S_{m_{\rm min}}$ and a maximum $S_{m_{\rm max}}$. We will assume that the measured flux at each of the catalogue positions will correspond to the flux density from one galaxy plus noise (i.e. we will neglect the possibility that more than one galaxy will dominate the flux in any given pixel). We can then bin the measured galaxy flux densities into flux density bins. Note that we will not consider in this analysis the possible angular correlations on the number density of galaxies. This extra information could further constrain the model, but this is usually only useful in the confusion noise limited case (the $P(D)$ analysis mentioned earlier). We also note that the ``detection'' image (i.e. the image from which the flux densities are extracted) will already implicitly contain this clustering information.

If we can calculate the probability, P$_i(k_i\,|\,dN/dS)$, of obtaining a given number of galaxies $k_i$ for bin $i$ given our proposed model $dN/dS$, then Bayes theorem states that the probability of the model given the data is, up to some normalisation, the probability of the data given the model:
\begin{equation}
P(dN/dS\,|\,k)\propto \prod_{i={\rm bin}} {\rm P}_i(k_i\,|\,dN/dS),
\end{equation}
where $\{k_i\}$ is our data.

\subsection{Probability function}
We will now calculate the theoretical expected distribution. Let us start by assuming that the number of galaxies in a given patch of sky obey a Poisson distribution with an average number per flux given by $\frac{dN}{dS}(S)$ (after integrating over the area), depending only on the flux $S$. Ignoring the effect of noise for the moment, the probability of observing $k_i$ galaxies over the observed sky area in the flux bin $i$ between $S_i$ and $S_i+\Delta S_i$ is given by 
\begin{equation}
P_i(k_i\,|\,dN/dS)=\frac{\lambda_i^{k_i} e^{-\lambda_i}}{k_i!}
\end{equation}
with $\lambda=\frac{dN}{dS}(S_i)\Delta S_i$ and $\frac{dN}{dS}(S_i)$ assumed constant across the interval $\Delta S_i$. For an infinitesimal interval $\Delta S_i=dS_i$ it should be enough to only consider the possibility of having zero galaxies in that bin, or at most one galaxy in each bin, for which the probability would be:
\begin{equation}
P_i(k_i=1\,|\,dN/dS)=\frac{\lambda_i e^{-\lambda_i}}{1!}\approx \lambda_i=\frac{dN}{dS}(S_i) dS_i.
\end{equation}

Adding noise, the measured flux $S_m$ for a given galaxy will be:
\begin{equation}
S_m=S+n
\end{equation}
where $S$ is the underlying "real" flux of the galaxy and $n$ the experimental noise which is assumed to obey a Gaussian distribution with average zero and variance $\sigma_n^2$. Taking again the case of an infinitesimal value, it is enough to consider the probability $P_m(S_m)$ of finding one single galaxy with measured flux in the interval $dS_m$, which is given by
\begin{equation}
P_m(S_m)dS_m\approx\int_{S_{min}}^{S_{max}} dS \frac{dN}{dS}(S) \frac{1}{\sigma_n\sqrt{2\pi}}\,e^{-\frac{1}{2}\left(\frac{S-S_m}{\sigma_n}\right)^2} dS_m.
\label{convolution}
\end{equation}
Here, $S_{\rm min}$ and $S_{\rm max}$ will be set by the model we choose, and $dN/dS$ is assumed to equal zero outside this range. The probability of finding $k$ galaxies in some larger, arbitrary range between $S_{m_i}$ and $S_{m_i}+\Delta S_{m_i}$, can be found by taking into account the different possibilities for distributing those $k_i$ galaxies in the interval:
\begin{equation}
P_i(k_i\,|\,dN/dS)=\frac{1}{k_i!}\left(\int_{Sm_i}^{Sm_i+\Delta Sm_i}P_m(S_m) dS_m\right)^{k_i} e^{-\int_{Sm_i}^{Sm_i+\Delta Sm_i}P_m(S_m) dSm}.
\label{PNeq}
\end{equation}
This is the salient equation of our method--with it, we can calculate the probability of any $dN/dS$ model reproducing a binned distribution from a set of measured, noise-dominated flux densities. Note that for a large number of galaxies this basically translates into a Gaussian distribution.

Similar forms of eq.~\ref{PNeq} can be found in \cite{Borys2003}, \cite{Laurent2005} and \cite{Coppin2006}, where the number counts of millimeter and sub-millimeter surveys are analysed. Note however that our application of the above equation is essentially different, since in the instrumental noise dominated regime we would not be capable of fitting for the number if the position of the galaxies was unknown.

Now, if $P_m$ was constant across the interval in the above equation, this would give
\begin{equation}
P_i(k_i\,|\,dN/dS)=\left(\Delta Sm_i P_m(S_i)\right)^{k_i}/{k_i}!\, e^{-\Delta Sm_i P_m(S_i)},
\end{equation}
which, for the zero noise case would fall back into the Poisson distribution:
\begin{equation}
P_i(k_i\,|\,dN/dS)=\left(\Delta Sm_i \frac{dN}{dS}(Sm_i)\right)^{k_i}/{k_i}!\, e^{-\Delta Sm_i \frac{dN}{dS}(Sm_i)}.
\end{equation}

\subsection{Simulation}
\label{simsec}
We will now simulate the expected distribution using a Monte-Carlo approach and compare it to the obtained function above. Note that the probability distribution is, of course, only a function of the model assumed, not of the data. We simulate the expected probability through the following steps:
\begin{indent}
\begin{enumerate}[1.]
\item Set a model parameterized as 
\begin{equation}
\frac{dN}{dS} = \begin{cases} C S^{\alpha} & \mbox{for } S_{min} < S < S_{max}\\ 0 & \text{Otherwise} \end{cases}
\end{equation}
\item Define an arbitrary set of bins which will be held constant throughout the process.
\item Generate a set of galaxies (fluxes) between $S_{min}$ and $S_{max}$ with the model above.
\item Add random noise to each flux assuming a Gaussian distribution for the noise.
\item Bin the resultant fluxes and record the number at each bin.
\item Repeat the process a large number of times, starting from step (3).
\end{enumerate}
\end{indent}
In this way, we can then obtain a histogram of the number of galaxies for each bin, which will give us a probability distribution for each bin.

\begin{figure*}
\includegraphics[scale=1.0]{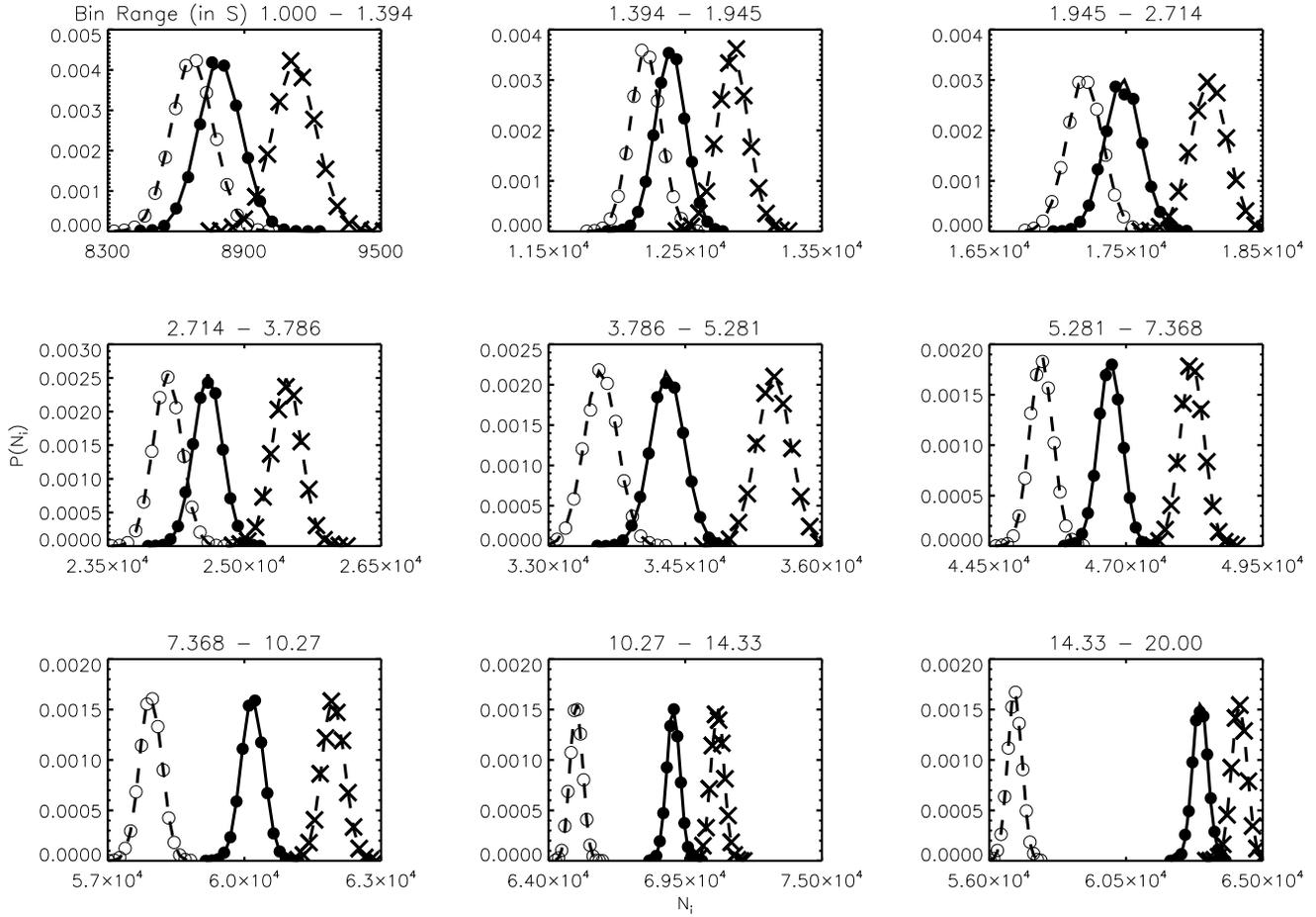}
\caption{Probability distributions for the number of sources in each bin. Curves represent the theoretical model and points are the results of our Monte Carlo simulations. The solid curve and filled circles have the same parameter values as those give in $\S$\ref{simsec}; the dashed line and open circles show the effect of decreasing $S_{max}$ by 25$\%$; dashed curves and crosses show the effect of decreasing $S_{min}$ by $5\%$.}
\label{PNsim}
\end{figure*}

We can now compare these results with eq. (\ref{PNeq}) above. For the example discussed here, we chose a model with $S_{min} = 1\, \mu$Jy, $S_{max} = 20\, \mu$Jy,  $dN/dS = 40\, S^{-1.50}$ Jy$^{-1}$ deg$^{-2}$, and $\sigma_n = 10\, \mu$Jy. The simulation was done over an area of 10 deg$^2$ and binned logarithmically between $1 < S_{\mu\text{Jy}} < 20$. We iterated through the steps $10^5$ times, resampling both $dN/dS$ and the noise distribution at each iteration, in order to generate probability distributions at each bin. We then calculated analytic values using eq.~\ref{PNeq} at each bin given the same model, bin sizes and sky area. This comparison can be seen in Fig.~\ref{PNsim}.

\section{Parameter Estimation Method}\label{sec:param_est}
We aim to constrain the four parameters, $S_{min}, S_{max}, C,$ and $\alpha$ of the $dN/dS$ model described above, using only the information from a noisy distribution. We assume that the form of the parameterization we choose can adequately model the number counts. If we denote the four unknown variables as $\theta$, the probability that a given set of model parameters correctly describes the data $k$ is:
\begin{equation}
P(\theta|k) = \frac{P(k|\theta)\,P(\theta)}{\int P(k|\theta)\,P(\theta)\,d\theta}
\end{equation}
where we are interested in the posterior distribution $P(\theta|k)$; $P(k|\theta)$ is the likelihood, $P(\theta)$ denotes the priors, and the denominator is the normalising Bayesian evidence term. In this analysis we ignore the Bayesian evidence term, since we are only solving for parameter values and not comparing between different models of $dN/dS$ parameterization. In this case, the Bayesian evidence term is a constant, (however, note that in the event that we want to perform model selection this Bayesian evidence term could be included). The likelihood $P(k|\theta)$ is calculated at each bin using eq.~(\ref{PNeq}). So all we need to do in order to constrain the four parameters is to sample from the parameter space, within some set of priors, and at each iteration over the parameter space compute the likelihood using the given set of observed, binned, noise dominated fluxes $k_i$. The total probability of the given set of model parameters is then just the product of the probability at each bin. 

\subsection{MCMC approach}
A direct gridding approach, where one generates four different grids spanning the prior range for each parameter and calculates the likelihood at each grid index, is effective but quickly becomes computationally inefficient as the grid resolutions are increased. An obvious solution to this is to implement a Markov chain Monte Carlo (MCMC) analysis. There is a weak degeneracy between the four variables we are solving for, so a standard Metropolis-Hastings MCMC approach has a tendency to ''get stuck'' in local maxima, as the degeneracy causes the target probability distribution to be multi-modal. This creates an inadequate sampling of the parameter space, and ultimately unconvergent or slowly convergent chains. We find that a reasonable solution to this is to employ a Parallel Tempering (PT) MCMC algorithm, as discussed in \citet{Liu:2001}.

The PT algorithm is similar to the Metropolis-Hastings algorithm in that we use the same acceptance/rejection criteria, with eq.~(\ref{PNeq}) as our likelihood function, but differs because the previous parameter state is not necessarily from the last iteration of the chain in question. The PT algorithm is basically just a way to swap parameter states between a ladder of chains that are running in parallel, where each chain has a different ``temperature'' $T$; we define the parameter $\beta$ = 1/$T$. The probability distributions for the chains are equal to $P(k|\theta)^{\beta} P(\theta)$. The first chain has $\beta = 1.0$ (the target distribution), with $\beta$ decreasing as you move down the ladder of chains, effectively broadening the probability distribution at each decrease in $\beta$. In our analysis here, we ran six chains in parallel with linearly decreasing $\beta$ values ranging between 1.0 and 10$^{-3}$. The chains to be swapped were chosen by generating a random uniform number $U$ between 1 and 5, where we propose to swap parameter states from chains $U$ and $U+1$. For chains $i$ and $i+1$, the probability that a swap occurs is equal to
\begin{equation}
r = \text{min}\left\{1,\,\frac{P(\theta_{i+1} | \,\beta_i)\,\,P(\theta_i |\, \beta_{i+1})}
{P(\theta_i |\, \beta_i)\,\,P(\theta_{i+1} |\, \beta_{i+1})}\right\}
\end{equation}
and a swap is carried out if $r$ is greater than or equal to a randomly generated uniform number.

With this method we were able to efficiently sample the entire parameter space of all four variables. In the end, each chain had a minimum length of $5\,\times 10^4$, and only results from the first chain are recorded. We executed the code many times, each time varying the initial values and the interval at which a parameter state is swapped. Parameter states between two chains were proposed to be swapped between every 30$^{\text{th}}$ and 10~000$^{\text{th}}$ iteration. (Note that in the limit of high swapping intervals, this method effectively becomes a Metropolis-Hastings algorithm ) In addition, we ran a number of simulations using the more standard Metropolis-Hastings algorithm. For our final analysis output, we mixed all the chains from both the PT and Metropolis-Hastings methods and set a burn-in of $5\,\times 10^3$ for each individual chain. Our final, mixed chains had lengths of $> 1.5\,\times 10^6$.

\subsection{Prior Selection} 
In general, MCMC solutions are directly dependent on the choice of priors. In our analysis we set conservative priors on $C$ and $S_{max}$, as we assume one will be able to gain ample insight from the detected population (i.e. sources with flux densities $\gtrsim 3 \,\sigma_n$). We also quantify the minimum flux density at which our method will work by relating $\sigma_{n}$ to the shot-noise component. 
Given that the integrated flux fluctuations from undetected galaxies should be below the instrumental noise, we impose the following condition on $S_{min}$ when we choose our initial parameter guesses to implement the MCMC:
\begin{equation}
  \sqrt{\Delta\Omega\int_{S_{min}}^{S_{max}}\,dS \frac{dN}{dS} \,\,S^{2}}\, < \eta\,\sigma_{n}
\end{equation}
where $\Delta\Omega$ is the sky area subtended by one pixel and $\eta$ is an arbitrarily small number less than 1. Thus, as the resolution of the telescope increases, the minimum flux at which our analysis will work decreases. Decreasing $\eta$ improves the errors on the model fitting; in our simulations, which are performed down to $S_{min} = 1~\mu$Jy with a resolution of 10$''$, we find an $\eta$ of $\sim 0.4$ is acceptable.

In addition to the aforementioned constraints, we also imposed uniform priors for each parameter separately: $-2.5 \le \alpha \le -0.1$; $0.01 \le S_{min,\mu\text{Jy}} \le 5.00$; $\sigma_{n} \le S_{max} \le 5\, \sigma_{n}$; $1 \le C_{\text{Jy}^{-1}\text{deg}^{-2}} \le 100$.

\begin{table*}
\begin{center}
\caption{Returned model parameters from simulations over various sky areas (noise rms of 10 $\mu$Jy). Errors are quoted at a 68\% confidence level.}
\renewcommand{\tabcolsep}{.3cm}
\begin{tabular}{c c c c c c c}
\hline
\hline\\[-2ex]
& Area (deg$^2$) &$\langle N_{tot} \rangle$ & $\alpha$ & $C$ (Jy$^{-1}$ deg$^{-2}$) & $S_{min}$ ($\mu$Jy) & $S_{max}$ ($\mu$Jy)  \\
\hline
&1 &$3.37\times 10^4$ & $-1.43^{+0.03}_{-0.09}$ & $46^{+32}_{-36}$ & $0.97^{+0.12}_{-0.16}$ & $20.10^{+0.80}_{-1.00}$ \\[1.ex]
&10 &$3.37\times 10^5$ & $-1.49^{+0.05}_{-0.07}$ & $45^{+32}_{-34}$ & $0.97^{+0.11}_{-0.12}$ & $20.00^{+0.40}_{-0.50}$ \\[1.ex]
&100 &$3.37\times 10^6$ & $-1.52^{+0.04}_{-0.06}$ & $34^{+19}_{-20}$ & $1.03^{+0.08}_{-0.10}$ & $20.00^{+0.20}_{-0.30}$ \\[1.ex]
\hline
Target: & & & $-1.50$ & $40 $ & 1.00 & 20.00 \\[1.ex]
\hline
\end{tabular}
\label{restable}
\end{center}
\end{table*}

\subsection{Simulation Results}

\begin{figure*}
\includegraphics[scale=.83]{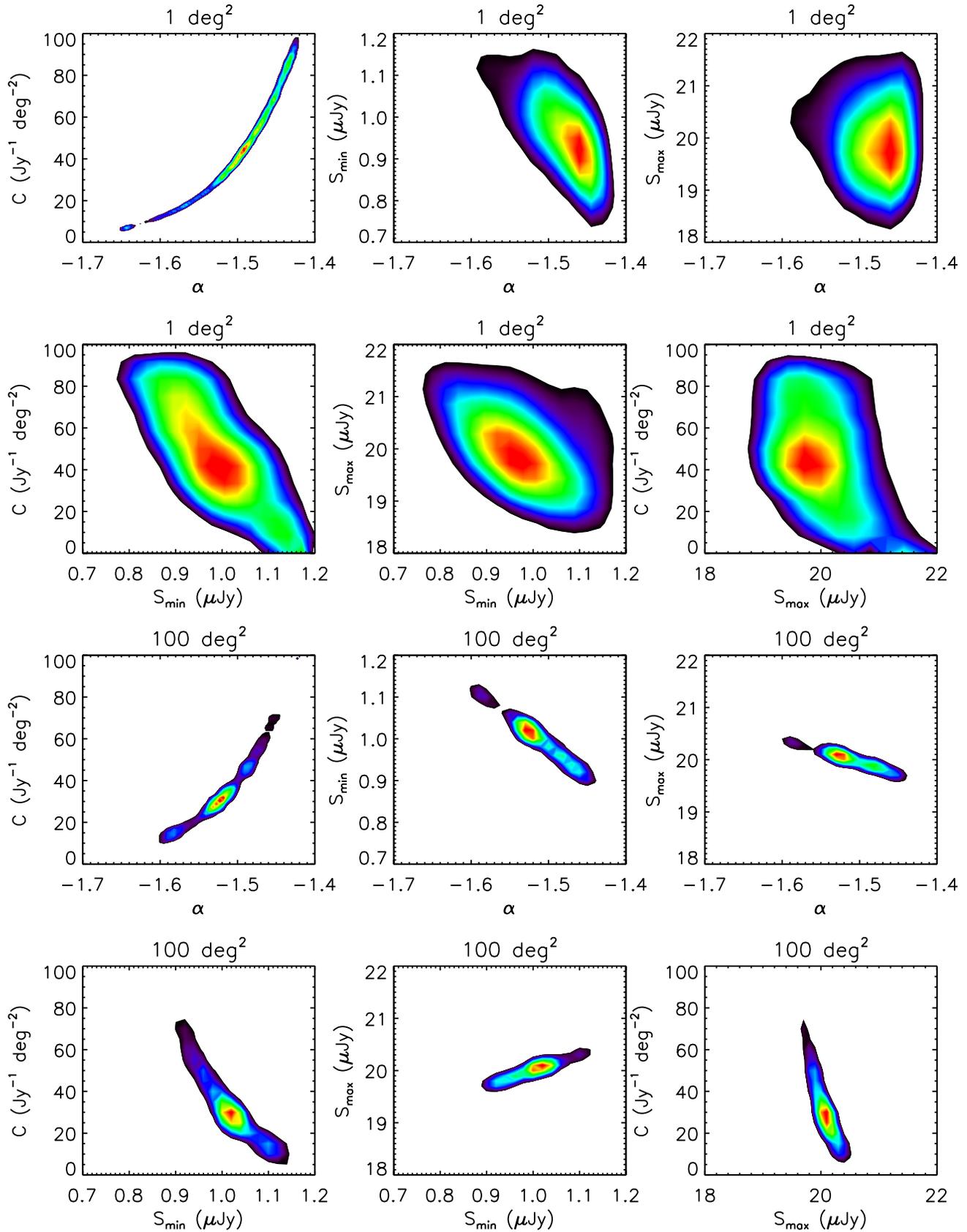}
\caption{68\% density plots for the model parameters, generated from the output chains of simulations over two different sky areas, as labeled in the title of each plot. 1 and 100 deg$^2$ respectively represent simulations done with $\langle\,N_{tot}\,\rangle \approx$ 10$^4$ and 10$^6$ sources, as can be seen in Table~\ref{restable}. Each contour encloses 68\% of the samples at its outer boundary.}
\label{simcont}
\end{figure*}

We test the efficacy of our method with a PT MCMC implementation by performing simulations over various sky areas with a constant $dN/dS$ model, effectively increasing the average total number of sources used for each increase in sky area. We simulated a noisy distribution parameterized in the same way as discussed in $\S$\ref{simsec}, used 10 bins logarithmically spaced between 1.0 and 50.0 $\mu$Jy, and imposed the prior constraints as discussed in the preceding section. With $\sim\,3\times 10^{4}$ simulated, noise-dominated sources (1~deg$^2$ with our chosen $dN/dS$ model), we are able to recover all four of the model parameters, within a confidence interval of 68\%. As we increase the number of sources to $\sim\,10^{6}$ (100~deg$^2$), the errors on the marginal posterior distributions fall by roughly 1/3; the best fit values for this many sources varies from the true model parameters by at most 7\% (e.g. $C$ in Table~\ref{restable}). Results of our simulations can be seen in Table~\ref{restable} and Fig.~\ref{simcont}. These results demonstrate the potential of our method under ideal conditions; next we demonstrate our method using real, noisy radio data.

\section{Analysis of Faint Radio Data}
We now turn to the application of our method to a real data set. In order to do this, we need noisy radio data with source flux densities extending well below the 5 $\sigma$ level. We obtained a noisy distribution of sources by extracting peak fluxes from a shallow radio image using the positions from a significantly deeper catalogue. This was done using data from the 2 deg$^2$ Cosmological Evolution Survey (COSMOS; \citealt{Scoville:2007}) and The Faint Images of the Radio Sky at Twenty centimeters survey (FIRST; \citealt{Becker:1994}).

FIRST is a large-area survey using the NRAO\footnote{The National Radio Astronomy Observatory is operated by Associated Universities, Inc., under cooperative agreement with the National Science Foundation.} Very Large Array (VLA). We used FIRST tiled images \citep{Becker:1995ei} covering the COSMOS field, which have a typical rms of 150 $\mu$Jy beam$^{-1}$, 5$''$ resolution, and a source detection threshold of 1~mJy. We extracted fluxes from this map using positions from observations done with the VLA-COSMOS survey, which is $\sim$ 13 times deeper than the FIRST survey.

The VLA-COSMOS Survey, centered on the COSMOS field, is comprised of both deep (2.5$''$ resolution) and large (1.5$''$ resolution) VLA projects at 1.4 GHz. We utilised the Large Project Catalogue \citep{Schinnerer2007}, which was updated and revised to account for bandwidth smearing \citep{Bondi:2008}. The catalogue contains only sources with peak flux densities above 5~$\sigma$, and 1~$\sigma_{n} \approx$~12~$\mu$Jy~beam$^{-1}$.

\begin{figure}
\begin{center}
\includegraphics[scale=.6]{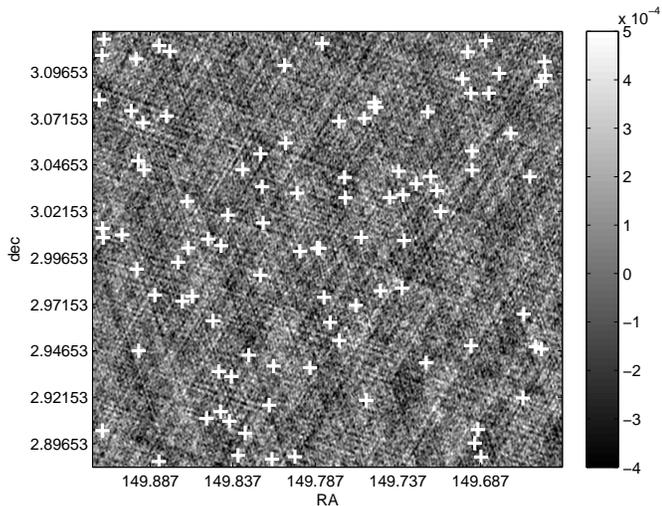}
\end{center}
\caption{A small portion of the FIRST image, in units of Jy/beam, with white crosses indicating the positions obtained from the deeper VLA-COSMOS survey. In this cropped image, none of the VLA-COSMOS positions are detections in the FIRST image.}
\label{fig:crop}
\end{figure}

\begin{figure}
\begin{center}
\includegraphics[scale=0.5]{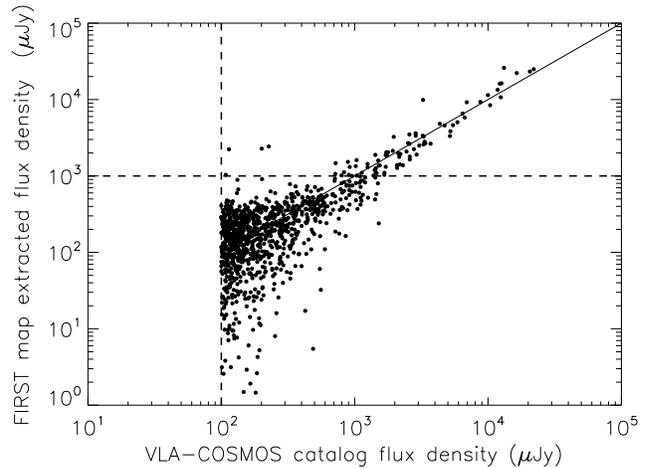}
\caption{VLA-COSMOS fluxes against their corresponding FIRST fluxes, as extracted from the FIRST map using VLA-COSMOS positions. The solid line indicates where both axes are equal and the dashed lines denote the detection limits for either survey. Note that the majority (97\%) of the FIRST fluxes are below the detection threshold of 1 mJy.}
\label{fig:fluxes}
\end{center}
\end{figure}

An example of non-detections in the FIRST image can be seen in Fig. ~\ref{fig:crop}. Fig.~\ref{fig:fluxes} shows the relationship between the 5 $\sigma$ fluxes from the VLA-COSMOS merged catalogue against those extracted from the FIRST map using the VLA-COSMOS positional information. Clearly the FIRST fluxes are noise dominated, with values extending down to a minimum of $-274\,\mu$Jy.

\begin{figure*}
\includegraphics[scale=0.8]{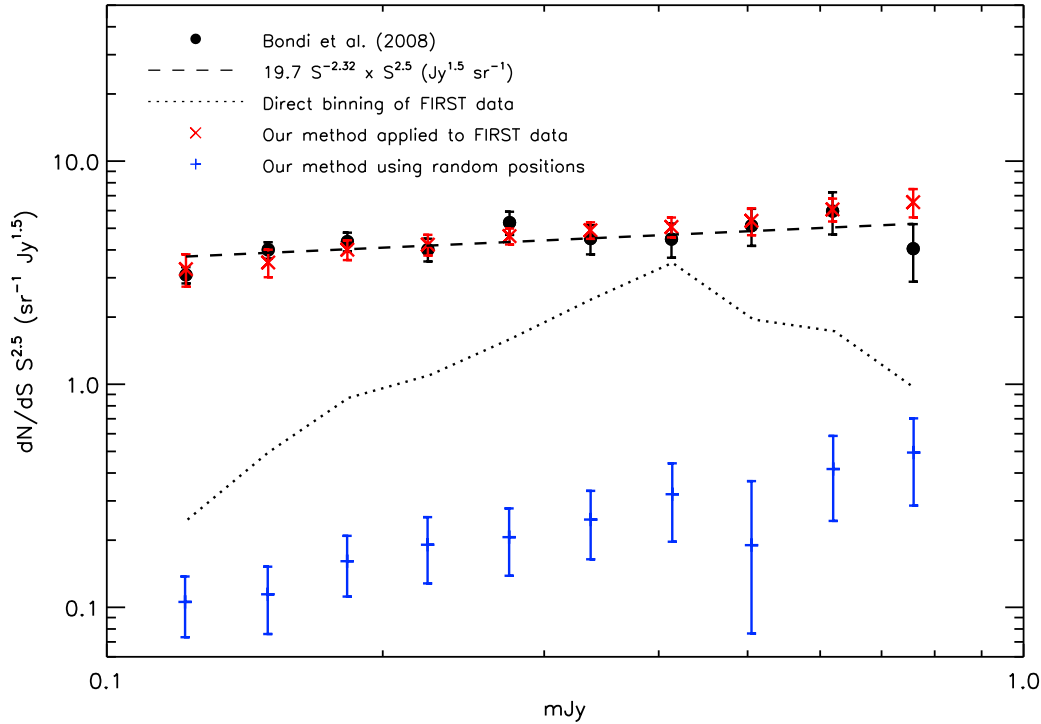}
\caption{Source counts for FIRST data below the detection threshold. The dotted line shows the raw FIRST flux densities extracted at the 5~$\sigma$ detected VLA-COSMOS positions. The dashed line shows the best fit to the $dN/dS$ analysis done in \citet{Bondi:2008} (black filled circles) according to our parameterization. Red crosses show the result of our method applied directly to the raw FIRST extracted flux densities. Blue crosses show the results of running our method using fluxes from the FIRST map taken at random positions. The blue and red error bars were calculated by generating histograms of $dN/dS$ for each $\alpha$ and $C$ pair in their respective MCMC output chains; errors are quoted at 68\%.}
\label{fig:bondifit}
\end{figure*}

\begin{table*}
\begin{center}
\caption{Best fit parameter values and errors are derived from marginal posterior distributions. The first row  shows the returned parameters from a blind extraction on the FIRST map, and the second row shows the returned model parameters for the noise-dominated FIRST fluxes as extracted at known positions. The last row shows our parameterization of the \citet{Bondi:2008} 5~$\sigma$ number counts.}
\renewcommand{\tabcolsep}{.3cm}
\begin{tabular}{c c c c c c c}
\hline
\hline\\[-2ex]
& Area (deg$^2$) &$N_{tot}$ & $\alpha$ & $C$ (Jy$^{-1}$ sr$^{-1}$) & $S_{min}$ ($\mu$Jy) & $S_{max}$ ($\mu$Jy)  \\
\hline
Random Positions: & 2.16 & 506  & $-1.66^{+0.05}_{-0.21}$ &$4.5^{+42.0}_{-3.0}$ & $1.03^{+1.40}_{-0.00}$ & $872.5^{+100.0}_{-85.0} $\\[1.ex]
FIRST:& 2.16 & $506$ & $-2.11^{+0.13}_{-0.14}$ & $ 7.2^{+28.8}_{-4.8}$ &$104.5^{+7.0}_{-8.0}$ & $715.0^{+50.0}_{-10.0}$ \\[1.ex] 
\hline
Target: & 2.16 & & $-2.32\pm{0.08}$ & $19.7_{-10.1}^{+38.6}$ & $110$ & $839$ \\[1.ex]
\hline 
\end{tabular}
\label{vlarestable}
\end{center}
\end{table*}

\subsection{FIRST Map Results}
\citet{Bondi:2008} performed a detailed analysis on the number counts properties of the sources in the VLA-COSMOS large project catalogue. In their analysis they parameterized the number counts as a sixth-order polynomial, so in order to include their results in our work, we fit their $dN/dS$ values to a function of the form $dN/dS = C S^{\alpha}$. For this fitting, we choose only the data with flux density values ranging between $ 110 \le S_{\mu\text{Jy}} \le 839$ as given in Table 3 of \citet{Bondi:2008}. The noise properties of the VLA-COSMOS maps used to generate the catalogues varies as a function of radius, with a 5 $\sigma_n$ detection limit at the edges of the map of $\sim 100\, \mu$Jy. An upper limit for the sources we use in this analysis is set at the 5 $\sigma_n$ detection limit of the FIRST survey (in \citealt{Becker:1995ei}, the FIRST team quoted $\gtrsim 6 \sigma_n$ as the detection limit for the FIRST maps). Our parameterized fit of the \citet{Bondi:2008} data agrees reasonably well, and can be seen in Figure~\ref{fig:bondifit}. We find the data can be parameterized as $dN/dS = 19.7^{+38.6}_{-10.1}\,S^{-2.32\pm 0.08}$ Jy$^{-1}$~sr$^{-1}$. We use these values to compare our PT MCMC analysis of the noise-dominated FIRST data.

We constrain the number counts properties of the FIRST data in a similar way as the simulations discussed in the previous section. We performed a PT MCMC analysis using the noise dominated FIRST extracted fluxes and set flat priors on the four unknown parameters we are solving for:  $-2.5 \le \alpha \le -0.1$; $1 \le S_{min,\mu\text{Jy}} \le 500$; $700 \le S_{max,\mu\text{Jy}} \le 1000$; $1 \le C_{\text{Sr}^{-1}\text{Jy}^{-1}} \le 100$. We performed the PT MCMC analysis to generate multiple chains, and mixed them to produce the results presented here. The analysis was done using 7 bins linearly spaced between $-270$ and $1000\,\mu$Jy. In Table~\ref{vlarestable} we show the best fit values as computed from each parameter's marginal posterior distribution, and Fig.~\ref{fig:bondifit} shows the best fit $dN/dS$ curve.

As a check of the results of our method using known source positions, we perform the analysis again using flux densities from a blind extraction on the FIRST image. We extract flux densities for the same number of galaxies as in the catalogue but at random pixel positions, and apply the same prior constraints, binning intervals and PT MCMC conditions. The marginal posterior distributions for each of the parameters returned from this analysis have larger errors, and the best-fit $dN/dS$ values are far from the expected values, as can be seen in figure \ref{fig:bondifit}.

\section{Simulations with Non-constant Noise Variance}
Throughout this paper we have considered $dN/dS$ distributions dominated by a Gaussian noise with a constant rms. In general, this is an ideal case since radio maps typically have a varying rms which primarily depends on the depth of the map--a function of multiple, non-uniform telescope pointings. In addition, bright sources will also affect the local rms of any given section of the map. Future generation radio surveys will be reduced with state of the art algorithms \cite[e.g.][]{Smirnov2011a,Smirnov2011b} which should greatly minimize non-uniformities, but small fluctuations on the instrumental noise rms are still expected. 

The most direct approach to deal with this changing rms will be to divide the sky into sections were we can take the rms to be very close to constant. If each section of sky is similar in size, so that $\langle\,N_{tot}\,\rangle$ is roughly constant, we can apply the previous described method to each section taking into account that the total probability of the model should be the product of the probability for each section. However, if the sections of sky vary in size, additional steps must be taken.

Accounting for varying sky sizes begins by binning the extracted fluxes in ``rms noise bins'' according to the rms noise associated with each extracted flux. This will introduce an extra step in the initial stage of our method: in addition to extracting noise-confused source flux densities at known positions, the rms noise of the area adjacent to each source must also be ``extracted''. This ``rms binning'' will allow one to perform the previous probability analysis for each group of galaxies which are approximated to have the same associated noise (i.e. each group of galaxies in the same rms bin). We can still assume a global model for the source counts, $dN/dS$, but the fitting to each rms noise bin $j$, with rms given by $\sigma_{n,j}$, should now be: $dN/dS_j=f_j\, dN/dS$, where $f_j$ is the fraction of the total number of galaxies that fall into that rms bin. The total probability for a given model $dN/dS$ is now the product of the probabilities for each noise bin $j$ with model $f_j\, dN/dS$. This is a straight-forward extension of the method outlined in this paper; given the range of $\sigma_{n}$ values in bin $j$, we can simply arrange the flux densities as a function of binned $\sigma_{n}$ values and calculate the probability in eq.~\ref{PNeq} for each $j$ bin increment of $\sigma_{n}$. We denote the flux density bins the same as before, with index $i$; we denote bin $j$ of $\sigma_{n}$ (i.e. $\sigma_{n,j}$) as just $\sigma_{j}$. The probability distribution of eq.~\ref{PNeq} then becomes a function of both bin index $i$ and rms noise index $j$, and contains the extra variable $f_j$:

\begin{equation}
\begin{split}
P_{i,j}(k_{i}\,|\,dN/dS,\sigma_{j}) = \frac{f_{j}}{k_{i}!}\left(\int_{Sm_i}^{Sm_i+\Delta Sm_i}P_m(S_m,\sigma_{j})\, dS_m\right)^{k_{i}} \times \\
\text{exp}\left(-\int_{Sm_i}^{Sm_i+\Delta Sm_i}P_m(S_m,\sigma_{j}) \,dS_m\right),
\end{split}
\end{equation}
and the total probability for any proposed $dN/dS$ distribution is
\begin{equation}
P(dN/dS,\sigma_{j}\,|\,k) \propto \prod_{j}\,\prod_{i}\,P_{i,j}(k_i,\,\sigma_{j}\,|\,dN/dS).
\end{equation}
Thus, with the additional step of binning the noise rms, one can extend the method presented in this paper to include the effects of a varying rms.

\section{Summary and Conclusions}\label{sec:conclusions}
The next generation of radio telescopes will survey huge areas of the sky with high sensitivity and resolution, so that statistical methods to analyse the data obtained will be increasingly important. Here we present a new method which is able to constrain the differential number counts, of specific source populations, down to flux density levels well below the detection threshold. In order to utilize this method, one needs to have knowledge of the positions of the noise-dominated flux densities in question, which will likely come from some auxiliary catalogues generated with data taken at other wavelengths. Our method does not allow holistic knowledge of the overall source population below the rms noise, but does allow one to build partial source counts of sources detected at other frequencies, using the flux densities extracted from noisy maps at known source positions. The method described in this paper models the $dN/dS$ distribution as a simple power law, but can be readily modified to include more complex $dN/dS$ models.

We demonstrate, with simulated populations parameterized by four values, what our method will be capable of with increasingly large numbers of sources. Our simulations aim to constrain the number counts between $0.1~\sigma$ and $2~\sigma$. With $\sim \,3 \times 10^4$ simulated, noise-dominated sources, we are able to recover all four of the $dN/dS$ model parameters within a 68\% confidence interval. Increasing the number of sources to $\sim\,10^6$, the best-fit values for our $dN/dS$ parameterization differ, at most, by only $7\%$.

Finally, we show a practical application of our method by performing the analysis on a set of noisy flux densities from the FIRST survey, extracted using well-known positions from the VLA-COSMOS survey, which is $\sim$~13 times deeper than FIRST. We use the results from a previous $dN/dS$ analysis done on the VLA-COSMOS data, which we find can be parameterized as $dN/dS = 19.7_{-10.1}^{+38.6}\times S^{-2.32\pm{0.08}}\,($sr$^{-1}$ Jy$^{-1})$. With our method applied to the noise-dominated FIRST data, we are able to recover the number counts from the noisy population alone as $dN/dS = 7.2^{+28.8}_{-4.8} \times S^{-2.11\pm{0.14}} \,($sr$^{-1}$ Jy$^{-1}$; $1~\sigma)$ with $110\,\mu$Jy~$\le S \le\,839~\mu$Jy. This analysis was performed using only 506 sources and extends below the rms of the FIRST survey.

\subsection*{Acknowledgements}
Support from the Science and Technology Foundation (FCT, Portugal) through the research grants PTDC/FIS-AST/2194/2012 (K.M.W., J.A., M.S,), PEst-OE/FIS/UI2751/2011 (K.M.W., J.A.), PTDC/CTE-AST/105287/2008 (K.M.W, J.A.), PTDC/FIS/100170/2008 (M.S.) and SFRH/BD/51791/2011 (K.M.W.) is gratefully acknowledged. The authors thankfully acknowledge the computer resources, technical expertise and assistance provided by CENTRA/IST. Computations were performed at the cluster ``Baltasar-Sete-S\'{o}is'' and supported by the DyBHo–256667 ERC Starting Grant.

\bibliographystyle{mn2e}
\bibliography{biblio}

\bsp

\label{lastpage}

\end{document}